\documentclass[symmetry,article,accept,pdftex,moreauthors]{Definitions/mdpi} 
\newcommand{\prd}{Phys.~Rev.~D}
\usepackage{booktabs,tabularx,multirow}
\usepackage{amsmath}  % 数学符号支持
\firstpage{1} 
\pubvolume{1}
\issuenum{1}
\articlenumber{0}
\pubyear{2025}
\copyrightyear{2025}
\externaleditor{Huanchen Hu} % More than 1 editor, please add `` and '' before the last editor name
\datereceived{27 December 2024} 
\daterevised{20 February 2025} % Comment out if no revised date
\dateaccepted{24 February 2025} 
\datepublished{ } 
\hreflink{https://doi.org/\vspace{-12pt}} % If needed use \linebreak
\Title{Long-Term Timing Analysis of PSR J1741$-$3016: 
 Efficient Noise Characterization Using PINT}
\TitleCitation{Long-Term Timing Analysis of PSR J1741$-$3016: Efficient Noise Characterization Using PINT}

\Author{Yirong Wen $^{1,2,3}$, Jingbo Wang $^{3,}$*,  Wenming Yan $^{1,4,}$*, Jianping Yuan $^{1,4}$, Na Wang $^{1,4}$, Yong Xia $^{1,2,3}$ and Jing Zou $^{1,2,3}$}

\AuthorNames{Yirong Wen, Jingbo Wang, Wenming Yan, Jianping Yuan, Na Wang, Yong Xia, and Jing Zou}

 \AuthorCitation{Y.R. Wen; 
 J.B. Wang; W.M. Yan; J.P. Yuan;  N. Wang; Y. Xia; J. Zou}

\address{%
$^{1}$ \quad Xinjiang Astronomical Observatories, Chinese Academy of Sciences, Urumqi 830011, 
 China;
\\
$^{2}$ \quad University of Chinese Academy of Sciences, Beijing 100049, China\\
$^{3}$ \quad Institute of Optoelectronic Technology, Lishui University, Lishui 323000, China \\
$^{4}$ \quad Xinjiang Key Laboratory of Radio Astrophysics, Urumqi 830011, China
}
\corres{Correspondence: 1983wangjingbo@163.com (J.W.); yanwm@xao.ac.cn (W.Y.)}

\abstract{The stable rotation of young pulsars is often interrupted by two non-deterministic phenomena: glitches and red timing noise. Timing noise provides insights into plasma and nuclear physics under extreme conditions.  The framework leverages rotational symmetry in pulsar spin-down models and temporal symmetry in noise processes to achieve computational efficiency, aligning with the journal’s focus on symmetry principles in physical systems. In this paper, we apply a novel frequentist framework developed within the PINT software package(v0.9.8) 
 to analyze single-pulsar noise processes. Using 17.5 years of pulse time-of-arrival (TOA) data for the young pulsar PSR J1741$-$3016, observed with the Nanshan 26 m radio telescope, we investigate its timing properties. In this study, we employed the Downhill Weighted Least-Squares Fitter to estimate the pulsar’s spin parameters and position. The Akaike Information Criterion (AIC) was used for model parameter selection. The results obtained with PINT were compared to those from {\sc ENTERPRISE} and {\sc TEMPONEST}, two Bayesian-based frameworks. We demonstrate that PINT achieves comparable results with significantly reduced computational costs. 
Additionally, the adequacy of the noise model can be readily verified through visual inspection tools.
Future research will utilize this framework to analyze timing noise across a large sample of young pulsars.}


\begin{document}

%%%%%%%%%%%%%%%%%%%%%%%%%%%%%%%%%%%%%%%%%%
\section{Introduction}

Pulsars, renowned for their extraordinary rotational stability, have long been utilized as cosmic laboratories for exploring a wide range of astrophysical phenomena. This is especially true for millisecond pulsars (MSPs), which exhibit rotation periods of just a few milliseconds, resulting from spin-up processes through accretion from companion stars \citep{2017JApA...38...42M}. Pulsar research has advanced our understanding of various domains, such as probing the dense matter equation of state \citep {2020NatAs...4...72C}, discovering planetary companions \citep{1992Natur.355..145W}, testing gravitational theories under strong-field conditions \citep {2021PhRvX..11d1050K}, and studying the properties of the interstellar medium \citep {2020A&A...644A.153D} and solar wind \citep {2021A&A...647A..84T}. Additionally, pulsar timing has contributed to the development of global time standards \citep{2020MNRAS.491.5951H} and refined solar system ephemerides \citep {2018MNRAS.481.5501C}. The method of pulsar timing, which involves tracking the arrival times of pulsar signals to monitor their rotational phase, has proven essential in recent breakthroughs, including the searching of a nanohertz gravitational wave background \citep{2023RAA....23g5024X} through Pulsar Timing Array (PTA) experiments \citep{Foster1990}.
%加个其他重要应用

Despite their precision as cosmic timekeepers, pulsars also exhibit irregularities known as ``timing noise'', which manifests as random deviations in pulse arrival times from a simple spin-down model. This noise is generally classified into two types: ``white'' noise and ``red'' noise. White noise is uniformly distributed across all frequencies and is typically associated with instrumental effects, radio frequency interference (RFI), or pulse shape fluctuations (e.g., pulse jitter) \citep{staelin1968}. In contrast, red noise is more prominent at low frequencies and is often linked to long-term processes, such as fluctuations in the interstellar medium density \citep{2013MNRAS.429.2161K} or the presence of nanohertz gravitational waves \citep{1983ApJ...265L..39H}. However, the primary source of red noise is believed to be intrinsic rotational instabilities within the pulsar itself, including ``glitch'' events, where the pulsar’s rotation rate suddenly accelerates, typically due to crustal stress release \citep{1969Natur.224..872B} or the unpinning of superfluid vortices \citep{1975Natur.256...25A, yuan2010}, as well as ``spin noise'', which manifests as long-term, red-spectrum fluctuations. While phenomena such as glitch recovery and changes in the pulsar's spin-down state may also play a role in generating red noise, the exact relationship between these phenomena and red noise remains under investigation \citep{hobbs2010, 2019MNRAS.489.3810P}.

The analysis of pulsar timing noise is fundamentally guided by symmetry principles.
Rotational symmetry underpins the deterministic spin-down model, while deviations from this
symmetry manifest as observable timing irregularities. Additionally, the scale-invariant
temporal correlations in red noise processes reflect a form of statistical symmetry across
observational timescales. By bridging frequentist and Bayesian approaches through
algorithmic symmetry, our methodology demonstrates how symmetry-driven analysis
enhances computational efficiency without compromising physical interpretability.

The accurate characterization of timing noise is critical for improving pulsar timing precision, as it impacts the utility of pulsars as time standards and their potential for detecting low-frequency gravitational waves.
%Extensive modeling of pulse arrival times reveals that phase residuals often exhibit erratic fluctuations in the pulsar’s rotation. 
 On shorter timescales, timing irregularities can often be modeled with low-order polynomials; 
 however, over long timescales, many pulsars exhibit significant deviations from these simple models. Notably, the braking index derived from a third-order polynomial fit is often much higher than what would be expected from magnetic dipole radiation, suggesting that the pulsar’s spin-down   is influenced by additional factors \citep{1984JApA....5..369B, 1988ApJ...326..947R}. 

In isolated pulsars, timing noise is primarily attributed to genuine changes in the rotation rate of the neutron star's crust rather than external processes affecting pulse emission or propagation \citep{1981ApJ...245.1060C}. Given the complexity of the neutron star’s internal structure and magnetosphere, multiple physical processes may contribute to the observed timing irregularities. All of these processes involve time-varying components of the torque acting on the pulsar’s crust. The two main sources of such torque are (i) an internal torque arising from the coupling between the crust and the superfluid interior, as observed in glitches, and (ii) an external ``radiation torque'' related to the pulsar’s magnetosphere. Although phenomena such as glitch recovery and changes in the pulsar's spin-down state may further affect the amplitude and characteristics of red noise, the precise relationship between these processes and red noise is still an ongoing area of study \citep{hobbs2010, 2019MNRAS.489.3810P}.

Pulsar timing involves the construction and incremental refinement of a timing model that aligns observed Times of Arrival (TOAs) with theoretical predictions, typically employing frequentist methods. This process is commonly performed using one of the three standard software packages: TEMPO 
 \citep{2015ascl.soft09002N}, TEMPO2 \citep{hobbs2006}, or PINT \citep{luo2021}. 
%These packages are often used interactively, allowing for adjustments to the timing model based on observed data. 
However, noise characterization is generally carried out separately in a Bayesian framework, with tools such as ENTERPRISE \citep{2024PhRvD.109j3012J} and TEMPONEST \citep{lentati2014} providing an estimation of noise parameters based on a post-fit timing model. ENTERPRISE, in particular, is also capable of modeling common deterministic and stochastic signals across multiple pulsars, such as the stochastic gravitational wave background and solar system ephemeris errors \citep{2023ApJ...952L..37A,2023RAA....23g5024X,2023ApJ...951L...6R}. The interdependence of timing and noise models requires them to be iteratively refined together, a process that is computationally expensive and time-consuming.

In contrast to these traditional approaches, PINT (Pulsar Timing Python Framework) offers a more efficient and flexible solution. Built on top of widely used scientific libraries, PINT was developed by the North American Nanohertz Observatory for Gravitational Waves (NANOGrav) \citep{2013ApJ...762...94D}. 
%This pure-Python framework allows for the integration of timing model refinement with noise characterization, enabling a more streamlined and modular approach to pulsar timing analysis.
PINT features a novel frequentist framework for noise characterization, allowing noise parameters to be simultaneously fitted with timing model parameters in a maximum-likelihood approach. This framework offers the ability to quickly obtain noise estimates and enables model comparison using the Akaike Information Criterion (AIC) \citep{burnham2004}, a tool not typically available in traditional Bayesian noise characterization methods. Furthermore, the frequentist approach in PINT can accelerate the iterative refinement of noise models during the initial stages of data preparation, providing a computationally efficient alternative to the more resource-intensive Bayesian approaches. Frequentist PINT-derived estimates provide an independent validation of the results of Bayesian models or serve as initial values for Markov Chain Monte Carlo (MCMC) samplers \citep{2022AnRSA...9..557J}, helping to reduce the time needed for convergence. In situations where Bayesian analysis is deemed computationally prohibitive, PINT provides a cost-effective alternative for noise characterization.

The structure of this paper is as follows:Section \ref{sec2} describes the observational dataset of PSR J1741$-$3016 and outlines the precision timing methodology implemented with the {\sc PINT} pulsar timing package. Section \ref{sec3}  presents the results of the timing analysis and model comparison. Finally, Section \ref{sec4} discusses and summarizes our results.

%%%%%%%%%%%%%%%%%%%%%%%%%%%%%%%%%%%%%%%%%%
\section{Observation and Data}\label{sec2}
PSR J1741$-$3016 was observed using the 25 m Nanshan radio telescope of the Xinjiang Astronomical Observatory, Chinese Academy of Sciences, located in Urumqi, China, spanning from August 2002 to December 2019, covering a total of 17 years and 4 months. 
The integration time for each observation varied between 4 and 16 min. The telescope's receiver operated across a frequency range of operating in L-band (1380-1700 MHz) with 320 MHz instantaneous bandwidth.

Before January 2010, the data were recorded using an analog filterbank (AFB) with 2 × 128 × 2.5 MHz channels \citep{wang2001}. From January 2010 onward, a digital filterbank (DFB) developed by the Australia Telescope National Facility (ATNF), replacing the AFB \citep{manchester2013}. This DFB was configured with 8-bit sampling and 1024 × 0.5 MHz channels, adequately covering the 320 MHz receiver bandwidth.The pulsar signals were processed with real-time folding algorithms, utilizing subintegration times of 1 minute for the Analog Filter Bank (AFB) system and 30 seconds for the Digital Filter Bank (DFB) system. %已重新修改这句话
The folded data were saved to disk with 256 phase bins per pulse period for the AFB data and 512 phase bins for the DFB data  \citep{yuan2017}.

PSR J1741$-$3016 was discovered in the Parkes multibeam pulsar survey. This survey covered the Galactic plane with $|b| < 5^\circ$ and $260^\circ < l < 50^\circ$ at a frequency of 1374 MHz and featured high sensitivity. A total of 370 new pulsars were discovered in this project, including PSR J1741$-$3016 \citep{morrisParkesMultibeamPulsar2002}.

The spin period of PSR J1741$-$3016 is approximately 0.255 s, with a period derivative ($\dot{P}$) of $8.99 \times 10^{-15}$
 yields a characteristic age of roughly $3.34 \times 10^5$ years.  The estimated distance to the pulsar is about 3.87 kpc. Additionally, the pulsar exhibits a GHz-Peaked Spectrum (GPS) with a peak frequency of approximately 620 MHz, based on both narrowband and wideband observations \citep{2021ApJ...922..125R}.

One of the primary motivations for selecting PSR J1741$-$3016 as the focus of this study is its extensive observational data and relatively uniform TOA uncertainties. Its prominent red noise characteristics make it an ideal candidate for investigating timing noise and long-term variability. By analyzing this pulsar, we aim to optimize and validate the timing algorithms and models in PINT, thereby enabling more accurate timing analyses for a large sample of pulsars in future studies.

The data reduction was carried out using the PSRCHIVE pulsar analysis software(v1.8) 
 \citep{hotan2004}. The initial steps involved removing radio frequency interference (RFI) and incoherently dedispersing the data using PSRCHIVE  \citep{straten2012}. This process combined frequency, time, and polarization channels to generate a mean pulse profile. Subsequently, a noise-free standard profile was constructed using the PAAS software package. 
 The mean pulse profiles for each observation were cross-correlated with the corresponding standard profile to extract topocentric pulse TOAs.

The initial parameters were obtained from the ATNF Pulsar Catalog \citep{2005AJ....129.1993M}. For preliminary processing, we removed outliers and bad data points. Each observed TOA was referenced to terrestrial time (TT), as realized by International Atomic Time (TAI), and subsequently converted to Barycentric Dynamical Time (TDB).
\section{Analysis and Results}\label{sec3}

Using the PINT software(v0.9.8), 
 the observed pulse arrival times were converted to times at the Solar System Barycenter (SSB) based on the DE421 solar system ephemeris \citep{luo2021}.
Subsequently, the TOAs at the SSB were fitted to the standard timing model for the pulse phase $N(t)$ as a function of time $t$:
\begin{equation}
N(t) = N_{\rm 0} + \nu (t - t_{\rm 0}) + \frac{1}{2} \dot{\nu} (t - t_{\rm 0})^2 + \frac{1}{6} \ddot{\nu} (t - t_{\rm 0})^3+\ldots
\label{eq:timmod}
\end{equation}
where 
 $N_0$ is the phase/pulse number at reference time $t_0$, and $\nu$, $\dot{\nu}$, and $\ddot{\nu}$ represent the spin frequency, its first derivative (spin-down rate), and second derivative, respectively.

Even in the absence of parameter degeneracies, the fitting algorithm can fail when strong nonlinear effects are present or when parameters approach their physical boundaries. Such nonlinearities necessitate robust fitting algorithms with regularized parameter updates, which allows for iterative adjustments that ensure the likelihood function remains well-defined throughout the fitting process. This iterative approach improves the fitting process in challenging cases where traditional methods may fail.

To address these challenges, we used the Downhill Weighted Least-Squares Fitter \texttt{DownhillWLSFitter} 
 in PINT \citep{susobhanan2024}, which significantly enhances the robustness of pulsar timing model fitting. Its advantage lies in employing a flexible update step \(b \rightarrow b - \lambda \hat{\beta}\), where \(\lambda \in (0, 1]\),  which allows for iterative adjustments that ensure the likelihood function remains well-defined throughout the fitting process. This iterative approach improves the fitting process in challenging cases where traditional methods may fail. 

\begin{figure}[H]
\includegraphics[width=10cm]{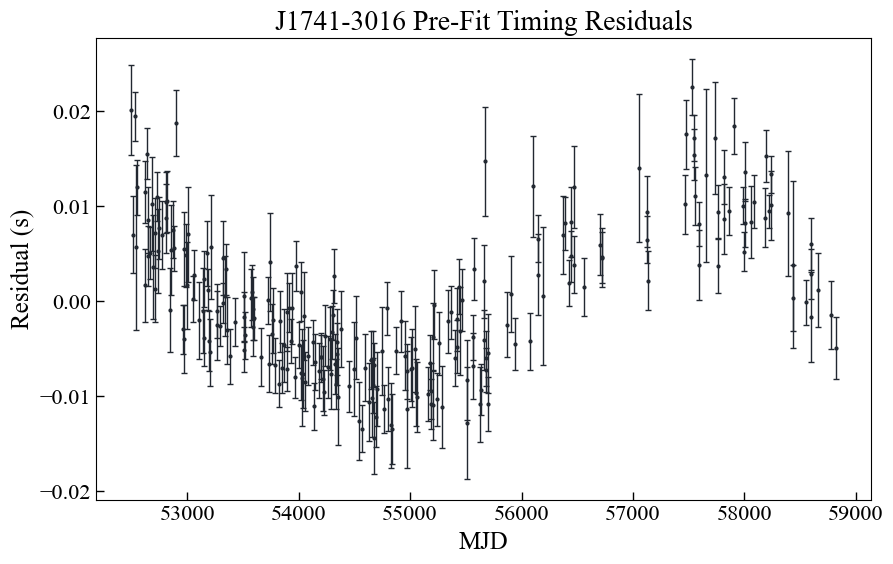}
\caption{J1741$-$3016 
 pre-fit timing residual.\label{fig1}}
\end{figure}

The fitted timing
 parameters obtained are presented in Table \ref{tab:observational}. This table compares the PSRCAT catalog values with our PINT and TEMPONEST results, showing improved positional accuracy (RA uncertainty reduced from 6 ms to 1 ms). The updated spin parameters ($\nu$, $\dot{\nu}$) align with previous studies but with tighter error bounds. The pre-fit residuals for PSR J1741$-$3016 are shown in Figure \ref{fig1}.

The timing residuals are defined as the differences between the observed pulse arrival times and predictions from the timing model as a function of Modified Julian Date (MJD) ({MJD = Julian Date $-$ 2,400,000.5, a continuous count of days primarily used in astronomical timing applications}). 
 Figure \ref{fig1} displays the pre-fit residuals spanning over 17 years, which exhibit two key features: (1) a long-term cubic trend indicating unmodeled spin frequency derivatives and (2) systematic offsets between different observational epochs (vertical groupings). These characteristics fundamentally motivated our subsequent red noise analysis using PINT's WaveX framework.

\vspace{-6pt}

\begin{table}[H]
\setlength{\tabcolsep}{3pt} % 压缩列间距
\small % 使用小字号
\caption{Observational characteristics estimates for PSR J1741$-$3016.}
\label{tab:observational}
\begin{adjustwidth}{-\extralength}{0cm} % 负边距扩展
\renewcommand{\arraystretch}{1.1} % 行高调整
\begin{tabular*}{\linewidth}{@{\extracolsep{\fill}} l c c c c l @{}}
\toprule
\multicolumn{1}{l}{\bfseries Parameter} & \multicolumn{3}{c}{\bfseries Measured Values} & \multicolumn{2}{c}{\bfseries Fitting Parameters} \\
\cmidrule{2-6}
 & \multicolumn{1}{c}{\textbf{PSRCAT}} & \multicolumn{1}{c}{\textbf{PINT}} & \multicolumn{1}{c}{\textbf{TEMPONEST}} & \multicolumn{1}{c}{\textbf{Parameter}} & \multicolumn{1}{c}{\textbf{Value}} \\
\midrule
RA, $\alpha$ (hh:mm:ss) $^{\mathrm{a}}$ & 17:41:07.04(6) & 17:41:06.89(1) & 17:41:07.04 & First TOA (MJD) & 52,495 \\
DEC.J, $\delta$ (dd:mm:ss) $^{\mathrm{b}}$ & $-$30:16:31(9) & $-$29:51:59.9(17) & $-$30:16:31 & Last TOA (MJD) & 58,819 \\
$\nu$ (s$^{-1}$) & 0.528053169233(8) & 0.528053169059(2) & 0.528053169360(2) & Timing epoch (MJD) & 55,665 \\
$\dot{\nu}$ (s$^{-2}$) & $-$2.51338(2) $\times$ $10^{-15}$ & $-$2.51682(2) $\times$ $10^{-15}$ & $-$2.51359(1) $\times$ $10^{-15}$ & Number of TOAs & 233 \\
RM (rad m$^{-2}$) $^{\mathrm{c}}$ & $-$450 & $-$450 & $-$450 & Solar system ephemeris model & DE421 \\
DM (cm$^{-3}$ pc) $^{\mathrm{d}}$ & 382 & 382 & 382 & Rms timing residual ($\upmu$s) & 8472.086 \\
\bottomrule
\end{tabular*}
\end{adjustwidth}
\noindent{\footnotesize{$^{\mathrm{a}}$ RA: right ascension; 
 $^{\mathrm{b}}$ DEC: declination; $^{\mathrm{c}}$ RM: rotation measure; $^{\mathrm{d}}$ DM: dispersion measure.}}
\end{table}

\subsection{Noise Model}
\subsubsection{White Noise}\label{sec3.1.1}

White noise refers to noise components that are independent for each TOA and can be modeled as an uncorrelated Gaussian noise process. It is characterized by a scale factor, which is utilized for correlation analysis in the \texttt{ScaleToaError} component module within PINT \citep{lentati2014}. Mathematically, white noise is represented by a diagonal matrix \(\boldsymbol{N}\), populated by the scaled TOA variances \(\varsigma_i^2\):
\begin{align}
    \varsigma_i^2 = F_i^2\left( \sigma_i^2 + Q_i^2 \right)\,,
\end{align}
where 
\begin{align}
    F_i &= \prod_a f_a^{\mathcal{F}_{ia}}\,,\\
    Q_i^2 &= \sum_a q_a^2 \mathcal{Q}_{ia}\,.
\end{align}
 where \(f_a\) and \(q_a\) are referred to as the EFAC (error factor) and EQUAD (error added in quadrature), respectively. %EE: Please check that the intended meaning has been retained
 Additionally, \(\mathcal{F}_{ia}\) and \(\mathcal{Q}_{ia}\) represent TOA selection masks, which can be 0 or 1 based on specific criteria that may depend on the observing epoch, observing frequency, observing system, and other relevant factors.
\subsubsection{Red Noise}

Red noise is typically associated with long-term irregularities in a pulsar's rotation, often referred to as spin noise or achromatic red noise (ARN). This type of noise arises from intrinsic pulsar dynamics, such as rotational distortions or irregular changes in spin. The power spectrum of achromatic red noise can be described using a Fourier Gaussian process model \citep{thenanogravcollaboration2015}:
\begin{align}
P(f) = A_{\rm red}^2\left( \frac{f}{f_{\rm yr}} \right)^{\gamma_{\rm red}}
\end{align}
where \( f \) denotes the frequency component of red noise, $A_{\rm red}$ is the amplitude of the red noise in $\upmu$s~${\rm yr}^{1/2}$, $\gamma_{\rm red}$ is the spectral index, and $f_{\rm yr} = 1{\rm yr}^{-1}$. This power-law model is implemented in the \texttt{PLRedNoise} component module within PINT.

When fitting red noise, PINT first fits the Fourier series as representation of achromatic red noise (\texttt{WaveX} component) and subsequently estimates the spectral parameters \citep{susobhanan2024}.

\subsection{Timing Parameters}
Table \ref{tab:observational} presents the newly obtained timing parameters for PSR J1741$-$3016 using the aforementioned fitting methods. This includes the updated position and spin parameters. The column labeled ``PSRCAT'' displays earlier position information and spin parameters obtained from the ATNF Pulsar Catalogue V2.5.1, along with the dispersion measure and Faraday rotation measure for PSR J1741$-$3016 \citep{manchester2013}. The position measurements in this study show a moderate improvement in precision compared to earlier results. %Additionally, PSR J1741$-$3016 exhibits improved timing measurements compared to those from version 1.54 of the ATNF Pulsar Catalogue. %Consequently, the resulting timing residuals for the pulsar analyzed reflect these enhancements in precision.这什么意思？改善可能来自于timing baseline 比较长？

\subsection{Timing Noise}  
\subsubsection{White Noise}  \label{sec3.3.1}

We characterized the white noise component of PSR J1741$-$3016 using EFAC and EQUAD parameters, as detailed in Section \ref{sec3.1.1}. The ECORR parameter, which typically accounts for correlated noise sources such as pulse jitter, radio frequency interference (RFI), polarization miscalibration, or interstellar scattering, 
which are correlated across different frequency sub-bands within the same observation. Since all our observations were conducted at around 1540 MHz, there was no need to include the ECORR parameter.

To evaluate the necessity of including the EFAC and EQUAD in the noise model, we employed the Akaike Information Criterion (AIC) in PINT for model comparison.

AIC, an asymptotically unbiased estimator of the expected Kullback--Leibler (K-L) information, is defined as:  
\begin{equation}  
\text{AIC} = 2q - 2\ln \hat{L}  
\end{equation}  
where \(q\) represents the total number of free parameters, including those from both the timing model and the noise model. \(L\) denotes the maximum likelihood of the model, and \( \hat{L} \) corresponds to the maximum likelihood value at optimal parameters. Among candidate models applied to the same data, the preferred model is the one that minimizes the AIC (with the smallest \( \text{AIC}_{\text{min}} \)) \citep{burnham2004}. Individual AIC results are not interpretable as they include arbitrary constants and are significantly influenced by sample size; therefore, we define the difference \(\Delta_i\):  
\begin{equation}  
\Delta_i = \text{AIC}_i - \text{AIC}_{\text{min}}  
\end{equation}  
where \(\text{AIC}_{\text{min}}\) represents the minimum among the \(R\) different \(\text{AIC}_i\) results (i.e., the minimum occurs at \(i = \text{min}\)). This transformation ensures that the best-fitting model has \(\Delta = 0\), while all other models yield positive \(\Delta_i\). Consequently, a larger \(\Delta_i\) indicates that the fitted model \(i\) is less likely to be the best approximating model within the candidate set.  

To evaluate the white noise model, we set up four parameter fitting scenarios:  
(1) fixed EFAC = 1 and EQUAD = 0;  
(2) free EFAC with EQUAD = 0;  
(3) fixed EFAC = 1 with free EQUAD;  
(4) free EFAC and EQUAD. 

Since observations were conducted with two distinct backends---an analog filterbank (AFB) before 2010 and a digital filterbank (DFB) after 2010---we included separate EFAC and EQUAD parameters for each. The Akaike Information Criterion (AIC) was used to select the optimal configuration. Table \ref{table:J1741-3016} presents the AIC results for these scenarios, showing that fitting the EFAC alone provided the best model.

The fitted EFAC values are as follows:

Since our data were obtained using two different backends, an analog filterbank (AFB) before 2010 and a digital filterbank (DFB) after 2010, we included two sets of EFAC and EQUAD parameters for each. The Akaike Information Criterion (AIC) was used to select the optimal configuration. Table \ref{table:J1741-3016} presents the AIC results for these scenarios, showing that fitting the EFAC alone provided the best model. Finally, after the fitting process, we obtained EFAC values are as follows: EFAC(FB) (Analog Filterbank backend, pre-2010) = 1.46 $\pm$ 0.08 and EFAC(Urum) (Urumqi Digital Filterbank backend, post-2010) = 2.65 $\pm$ 0.23.

\begin{table}[H]  
%\centering % 添加这一行以居中表格  
\caption{AIC differences for different noise model configurations for simulation with EFAC and EQUAD (white noise only).}  \setlength{\tabcolsep}{12.2mm}
\begin{tabular}{llc}
\toprule
\textbf{Parameter Combination} &\textbf{AIC} & \textbf{Difference}\\  
\midrule
EFAC = 1, EQUAD = 0 & $-$1867 &311\\  
EFAC = 1, EQUAD free & $-$2168 &10\\   
EFAC free, EQUAD = 0 & $-$2178 &0\\   
EFAC free, EQUAD free & $-$2173 &5\\  
\bottomrule 
\end{tabular}
\label{table:J1741-3016}  
\end{table}

\subsubsection{Red Noise}

\begin{figure}[H]% 居中对齐
\includegraphics[width=9cm]{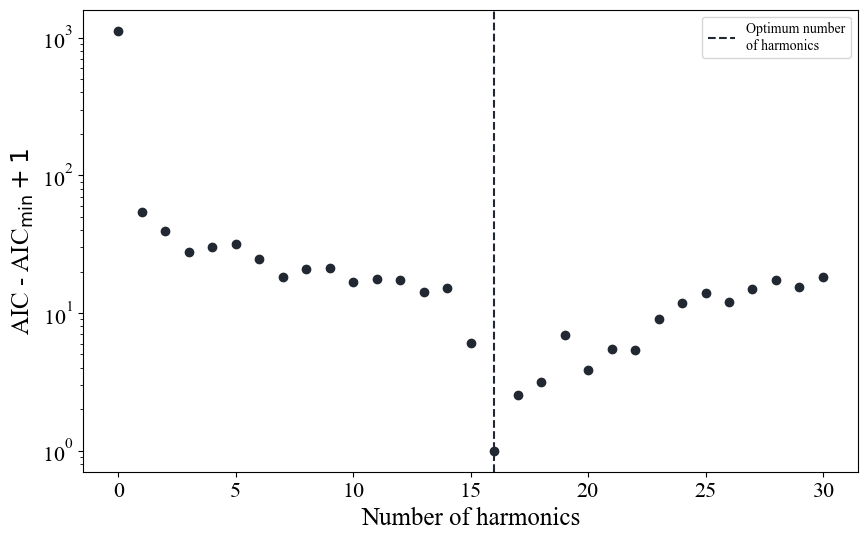} % 设置固定宽度
\caption{Variation in the Adjusted Akaike Information Criterion (AIC $-$ AIC$_{\text{min}}$ $+$ 1) as a function of the number of harmonics used in the model.\label{fig2}}
\end{figure}

\begin{figure}[H]% 居中对齐
\includegraphics[width=10cm]{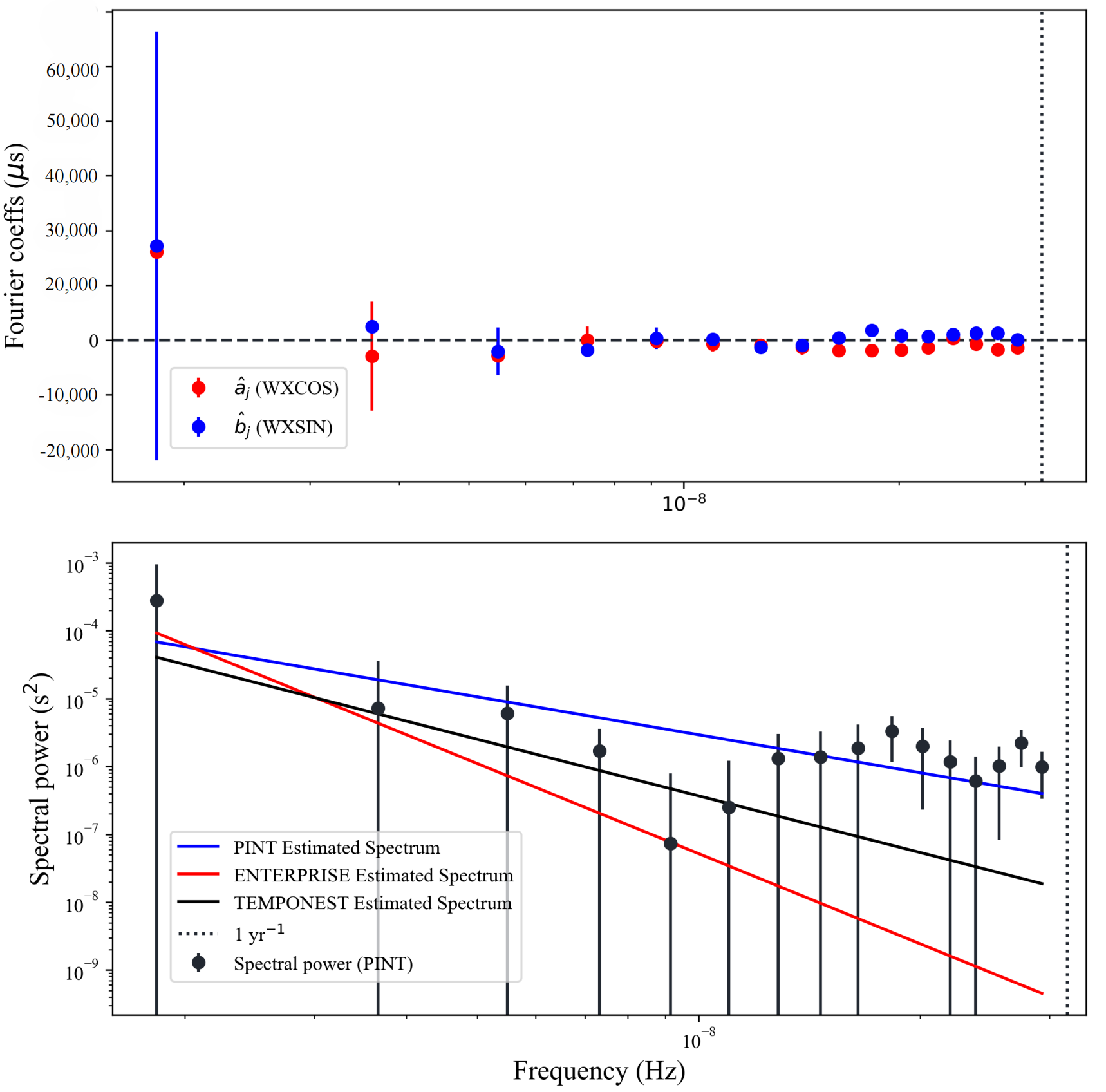} % 设置固定宽度
\caption{Harmonic parameter 
 estimation results for red noise and power-law fit analysis. The upper panel shows the estimates of the Fourier coefficients derived using the WaveX model with 16 harmonics, as determined by the minimum AIC difference. The vertical dashed line in both panels marks the reference frequency of $1\,\text{year}^{-1}$, and the horizontal dashed line in the upper panel indicates the zero-amplitude baseline. The lower panel presents the best-fit power-law model applied to these coefficients.\label{fig3}}
\end{figure}

We employed the WaveX model to characterize the noise, as described in Section~\ref{sec3.3.1}, utilizing harmonics at frequencies corresponding to the fundamental frequency of the span $T_{\text{span}}^{-1}$, where $T_{\text{span}}$ is the total observation span. To determine the optimal number of harmonics required for modeling the noise, we fitted the TOAs using varying numbers of harmonics and calculated the corresponding AIC for each case. 
Figure \ref{fig2} illustrates the AIC comparison results, showing that the minimum AIC difference is achieved with 16 harmonics. The upper panel of Figure \ref{fig3} displays the maximum likelihood estimates of the Fourier coefficients incorporated into the model. A power-law fit was applied to these estimated coefficients, and the lower panel of Figure \ref{fig3} shows the resulting best-fit power-law model.

\vspace{-3pt}
%-------------------------------------------------------------

\vspace{-6pt}

\begin{figure}[H]
% 添加居中对齐
\includegraphics[width=10cm]{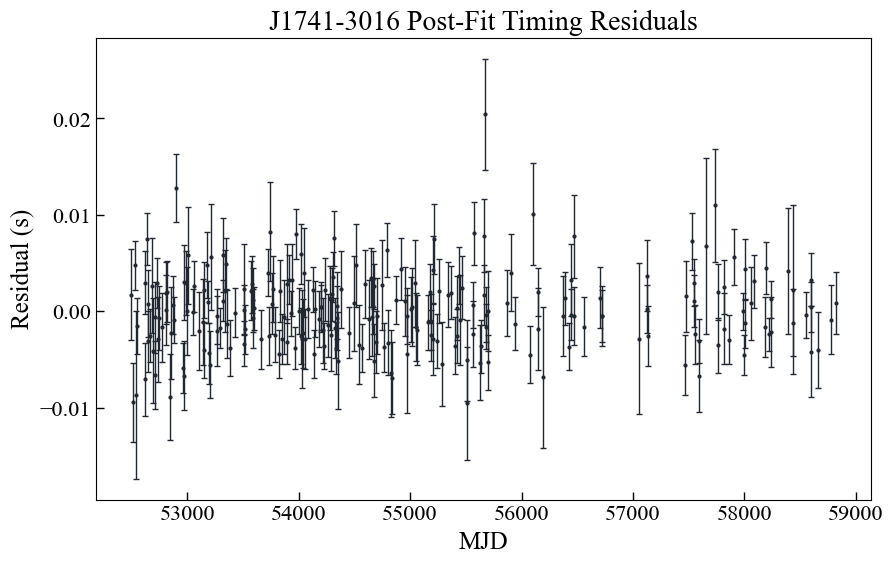} % 设置固定宽度
\caption{J1741$-$3016 post-fit timing residuals.\label{fig4}}
\end{figure}

The final fitting results of the red noise parameters are as follows: $\gamma = 1.86 \pm 0.37$ and $\log_{10} A = -9.07 \pm 0.13$. These results are further compared with Bayesian frameworks in Section 3.4 .
 The residuals after fitting are illustrated in Figure \ref{fig4}, revealing that the model provides a good fit, and almost only Gaussian random components remain.

\unskip

\subsection{Results Comparison} 
In the preceding sections,
 we detailed the modeling of timing noise for PSR J1741$-$3016, including the characterization of white and red noise components. Using the PINT framework, we achieved robust parameter estimation and model selection, with white noise results summarized in Table \ref{table:J1741-3016}. A detailed comparison of these findings with Bayesian frameworks ({\sc TEMPONEST} and {\sc ENTERPRISE}), including implications for red noise characterization, will be presented later in this section.

For {\sc TEMPONEST} and {\sc ENTERPRISE}, the  prior setting for white and red noise-related parameters from PINT are detailed in Table \ref{tab:3}. Similar prior settings have been widely used in the literature.
The absence of multiband observations posed a significant limitation, preventing a robust separation of potential dispersion measure (DM) variations from intrinsic red noise in the timing residuals of PSR J1741$-$3016. Consequently, advanced methodologies such as \texttt{DMWaveX}, which require high-precision, multi-frequency data to model and characterize DM fluctuations, could not be applied reliably in this analysis.
The lack of multiband observations precludes a robust disentanglement of potential dispersion measure (DM) variations from intrinsic red noise in the timing residuals of PSR J1741$-$3016. As a result, methodologies such as \texttt{DMWaveX}, which require high-precision, multi-frequency data to model and characterize DM fluctuations, could not be applied reliably in this analysis.

\begin{figure}[H]
\subfloat[\centering  
]{\includegraphics[width=6cm]{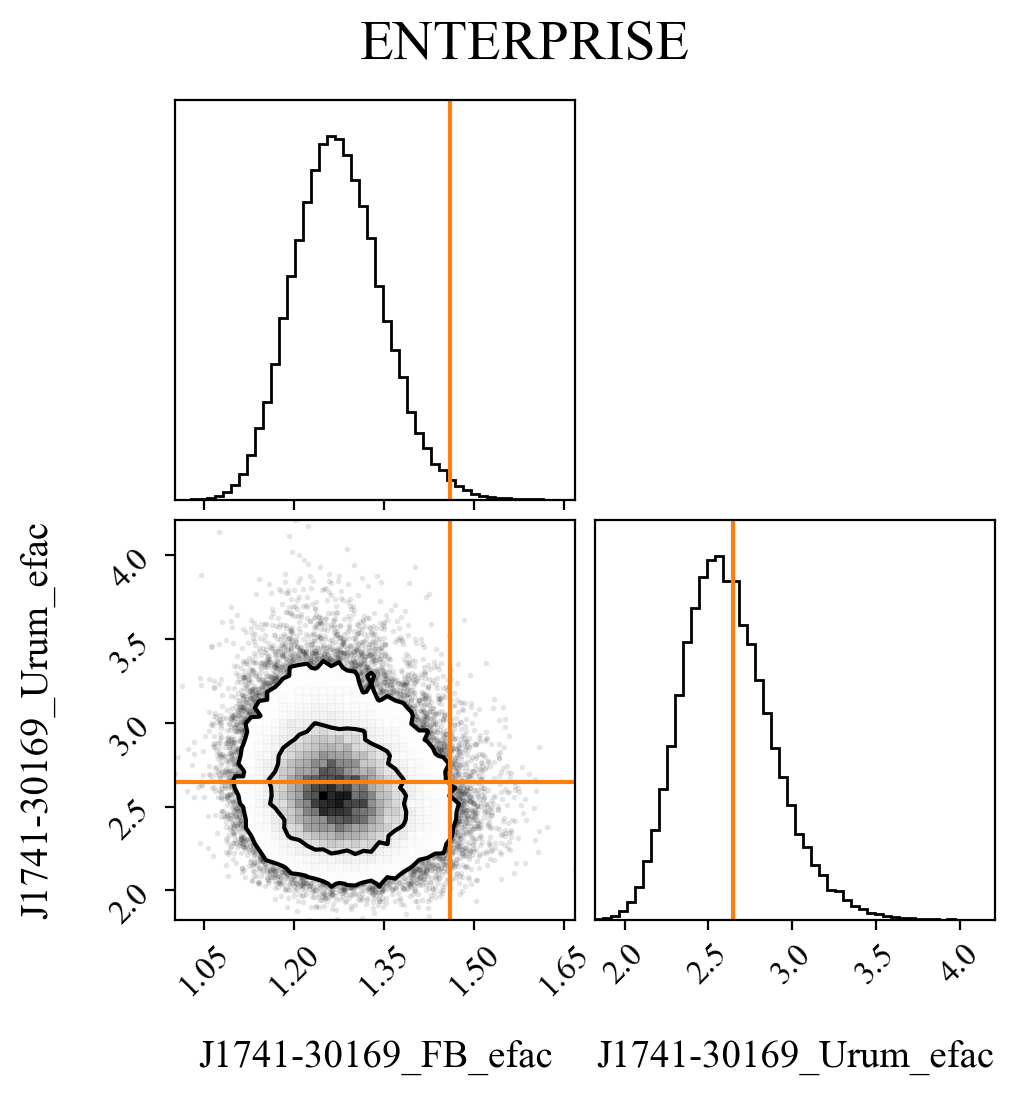}}
\hspace{0cm}  % 控制子图之间的水平间距
\subfloat[\centering  
]{\includegraphics[width=6cm]{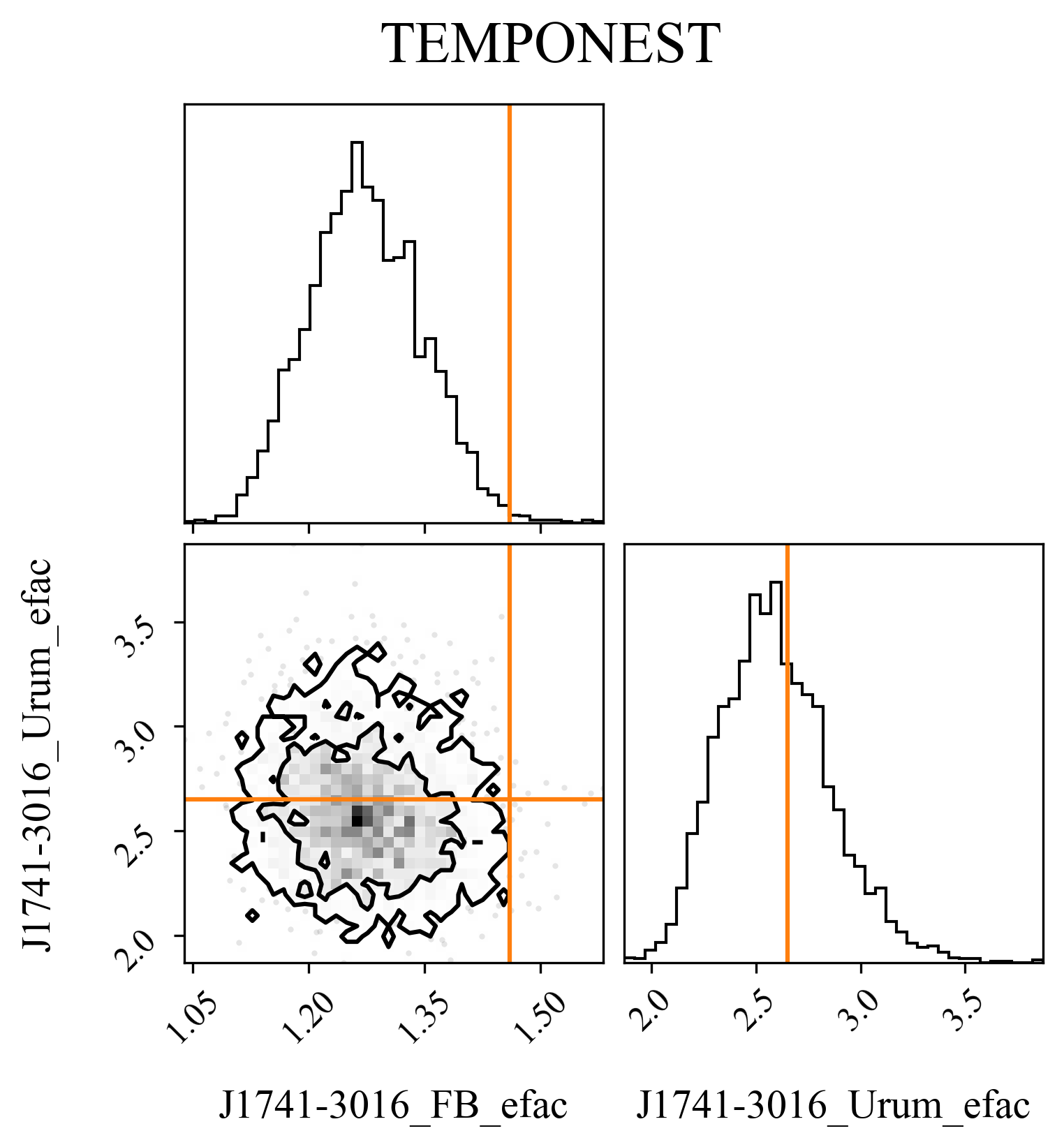}}
\caption{Comparison of white noise parameter corner plots from two different sampling methods: (\textbf{a}) ENTERPRISE and (\textbf{b}) TEMPONEST. The orange crosshairs indicate the PINT-derived values. The intersection coordinates correspond to the best-fit parameters obtained using the frequentist approach in PINT.
}\label{Fig:comparison}
\end{figure}

\begin{figure}[H]
\subfloat[\centering  
]{\includegraphics[width=6cm]{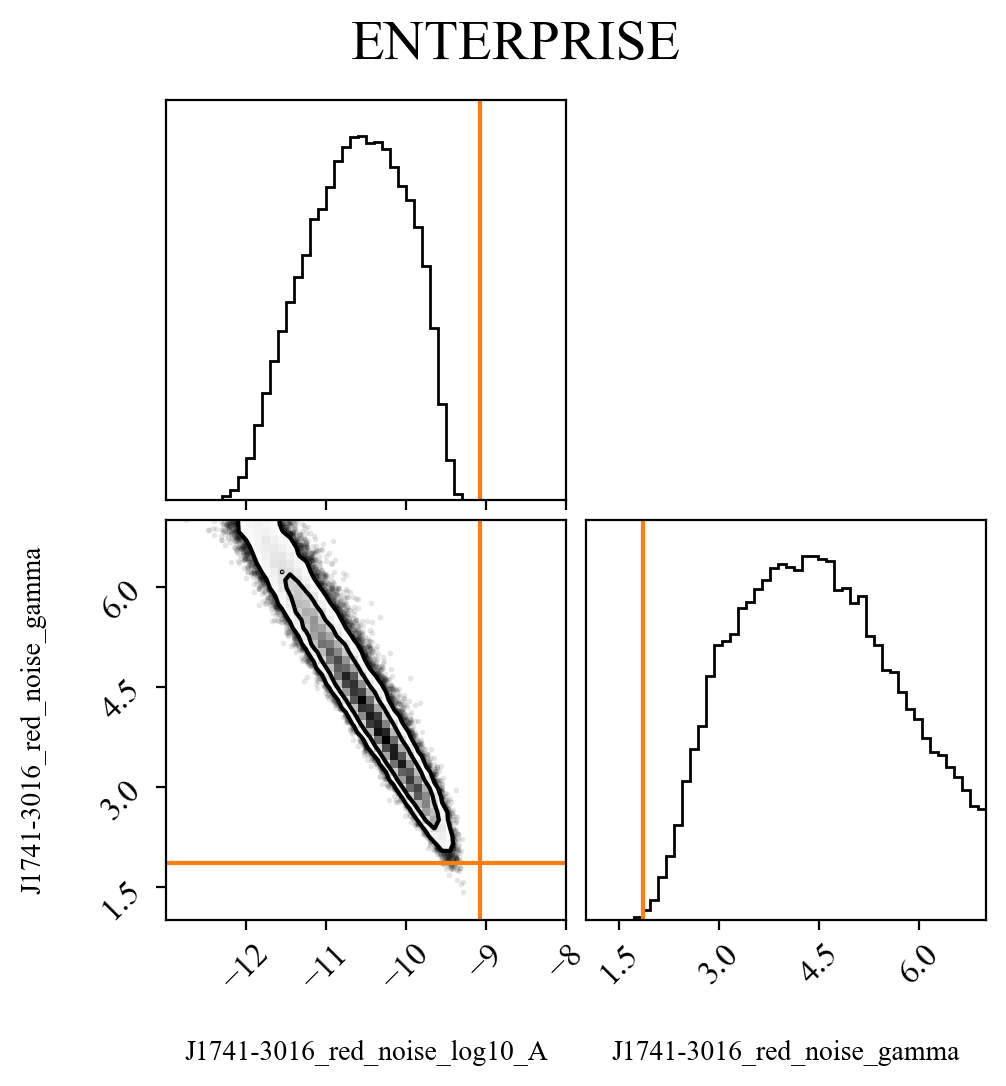}}
\hspace{0.5cm}  % 控制子图之间的水平间距
\subfloat[\centering  
]{\includegraphics[width=6cm]{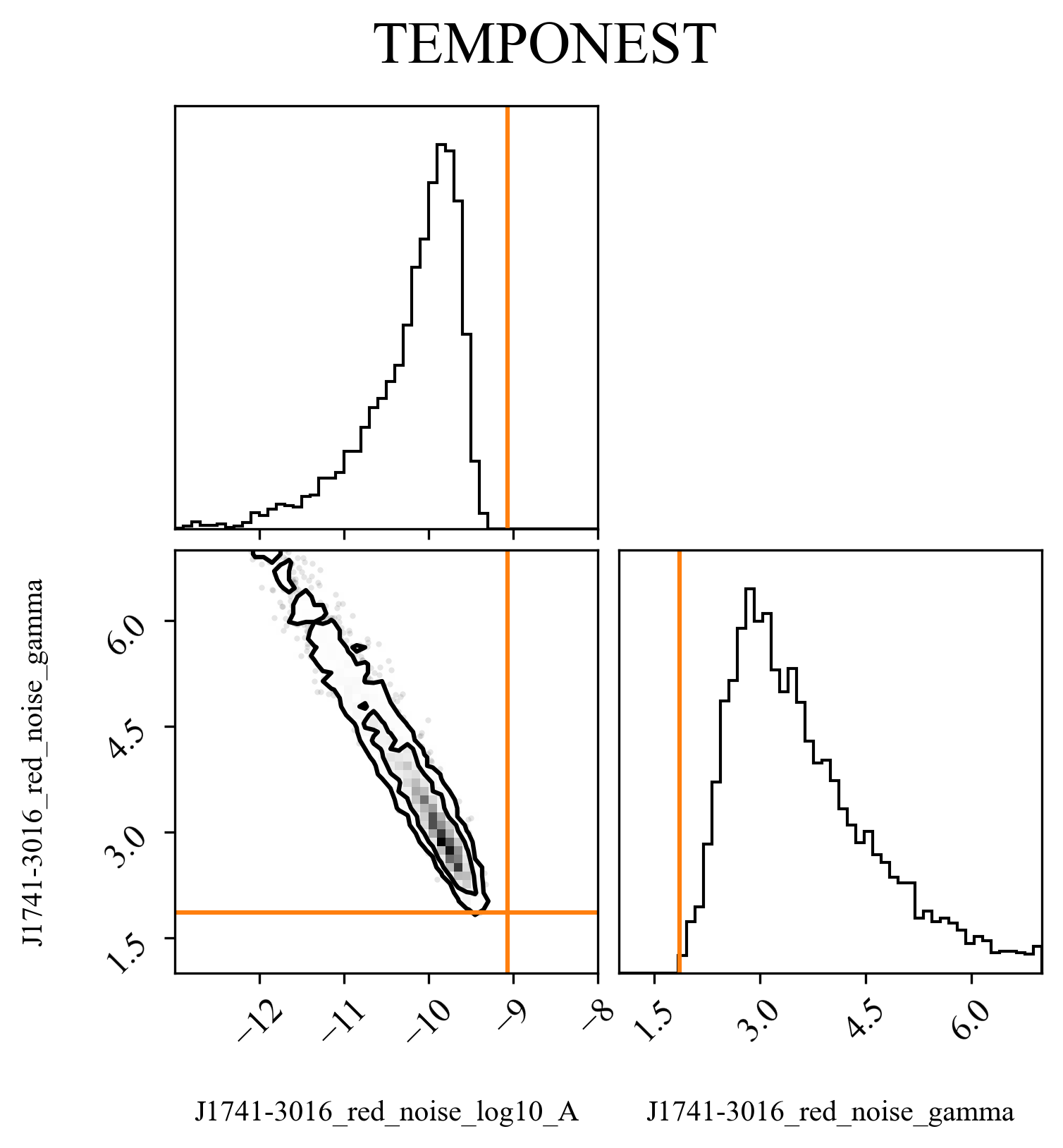}}
\caption{Comparison of power law spectral parameters using different methods: (\textbf{a}) PINT and ENTERPRISE and (\textbf{b}) PINT and TEMPONEST. The intersecting orange lines mark PINT-derived estimates, 
providing direct comparison with Bayesian posterior distributions.\label{Fig:comparison_power_law}}
\end{figure}

\begin{table}[H]
    \caption{Prior ranges on pulsar and timing noise parameters.}\setlength{\tabcolsep}{4.7mm}
    \label{tab:3}
    \begin{tabular}{llll}
        \toprule
        \textbf{Parameter} & \textbf{Symbol [units]} & \textbf{Prior Range} & \textbf{Prior Type} \\
        \midrule
%        Spin-frequency and derivatives & $\nu$, $\dot{\nu}$, $\ddot{\nu}$ [Hz, s$^{-2}$, s$^{-3}$]     & $\pm x^{\star}\times\Delta_{\mathrm{param}}$ & Uniform \\
        White noise fitting factor & EFAC                              & ($-$1, 2) & Uniform  \\
        Red noise amplitude & $A$ [yr$^{3/2}$]                  & ($-$20, $-3$) & log-Uniform  \\
        Red noise spectral index & $\gamma$                           & (0, 20)   & Uniform \\
       \bottomrule
    \end{tabular}
\end{table}
It is evident from Table \ref{tab:4} that the two methods yield consistent results for the white noise parameters. However,  {\sc TEMPONEST} shows larger uncertainties, as illustrated in the corner plots (Figure \ref{Fig:comparison}). Similarly, the results for the red noise parameters are also comparable across methods, though the convergence profile of ENTERPRISE appears to be more dispersed (Figure \ref{Fig:comparison_power_law}).

\vspace{-6pt}

\vspace{-18pt}
\vspace{-6pt}
\begin{table}[H]
    \caption{Timing 
 noise parameter comparison. }\setlength{\tabcolsep}{6.6mm}
    \label{tab:4}
    \begin{tabular}{llll}
        \toprule
        \textbf{Parameter} & \textbf{PINT} & \textbf{ENTERPRISE} & \textbf{TEMPONEST} \\
       \midrule

        EFAC (
FB) & 1.46 $\pm$ 0.08& 1.27 $\pm$ 0.14 & 1.27 $\pm$ 1.06 \\

        EFAC (Urum) & 2.65 $\pm$ 0.23& 2.59 $\pm$ 0.53 & 2.58 $\pm$ 1.11 \\
        $\gamma$ & 1.86 $\pm$ 0.37& 4.41$ \pm$ 2.13   & 3.72 $\pm$ 1.21  \\
        Log$_{10}$\textit{A} & $-$9.07 $\pm$ 0.13& $-$10.59 $\pm$ 1.09    & $-$9.75 $\pm$ 0.60 \\

       \bottomrule
    \end{tabular}

%    {$^{\star}x$ lies between $100 - 100000$ depending on the pulsar.}
\end{table}

When comparing the results of {\sc PINT} with those obtained using {\sc TEMPONEST} and {\sc ENTERPRISE}, we find a slight difference in the EFAC(FB), while the EFAC(Urum) fitting outcomes closely align  with those from the other two  Bayesian methods. Regarding the red noise fitting, there are minor differences in the fitted $\gamma$, although the fitted results for $\log_{10}A$ are relatively consistent.

%In terms of computation time, PINT employs a frequentist method based on maximum likelihood estimation, which is significantly faster compared to the Bayesian methods that require sampling to obtain parameter estimates. Thus, for a larger number of pulsars, opting for PINT proves to be a convenient and efficient approach.

In terms of computational efficiency, PINT demonstrates significant advantages over Bayesian frameworks. As shown in Table \ref{tab:5}, PINT(v0.9.8) completes the noise characterization process in 165.8 s with only 4\% average CPU utilization, compared to 347.8 s (26\% CPU) for ENTERPRISE(v3.4.2) and 283.2 s (34\% CPU) for TEMPONEST(v0.1.0). This efficiency gain stems from PINT's optimized frequentist approach that avoids computationally intensive Bayesian sampling while maintaining comparable accuracy.
\begin{table}[H]
\caption{Computational performance comparison. {All benchmarks 
 were performed on identical hardware: Dual Intel Xeon 6346 (3.1 GHz, 16C), 1 TB DDR4-3200 RAM, and NVIDIA RTX 3090 GPUs.}}\setlength{\tabcolsep}{9.2mm}
\label{tab:5}
\begin{tabularx}{\textwidth}{lcc}
\toprule
\textbf{Framework} & \textbf{Runtime (s)} & \textbf{CPU Utilization (\%)} \\
\midrule
ENTERPRISE & 347.8 & 26 \\
PINT & 165.8 & 4 \\
TEMPONEST & 283.2 & 34 \\
\bottomrule
\end{tabularx}
\end{table}

\section{Discussion and Summary} \label{sec4}

PSR J1741$-$3016 is a young radio pulsar with a characteristic age of $3.34 \times 10^5$ years. This study presented a comprehensive timing analysis of PSR J1741$-$3016, based on 17.5~years of observational data collected with the Nanshan 25 m radio telescope. Using the \textsc{PINT} framework, we modeled the pulsar's spin parameters, position, and timing noise, including both white noise and red noise components. The Akaike Information Criterion (AIC) was applied to evaluate various noise models and parameter configurations, achieving an optimal balance between model complexity and goodness of fit.   This highlights PINT's capability to handle datasets effectively while maintaining accurate noise characterization.

The red noise modeling process also revealed differences in the number of Fourier coefficients employed by each framework: PINT required 16 coefficients, TEMPONEST utilized 52, and ENTERPRISE applied the default 30 coefficients. These variations reflect differing approaches to noise modeling, with PINT striking a balance between computational efficiency and model complexity. This highlights PINT's capability to handle large datasets effectively while maintaining accurate noise characterization.

One notable outcome of this analysis is the overall consistency of the timing solutions obtained using \textsc{PINT}, \textsc{TEMPONEST}, and \textsc{ENTERPRISE}. While minor differences were observed in specific parameters, particularly the red noise spectral indices, the general agreement among these methods underscores the robustness of the frameworks. Notably, these models demonstrated exceptional efficiency, delivering comparable results in significantly shorter computation times, making them practical tools for timing studies of young pulsars. %EE: Please check that the intended meaning has been retained
 
%EE: Repeated text deleted

The advanced visualization tools in \textsc{PINT}, including Fourier coefficient plots, power spectrum estimates, power-law fits, and post-fit residuals further illustrate its strengths. These features enabled us to assess the appropriateness and reasonableness of the red noise model. For the red noise parameters, we obtain $\log_{10} A_{\text{red}} = -9.07 \pm 0.13$ and $\gamma = 1.86 \pm 0.37$. Furthermore, our results are consistent with the conclusion that the strength of timing noise scales proportionally to $\nu^{1} |\dot{\nu}|^{-0.6 \pm 0.1}$ \cite{2019MNRAS.489.3810P}, where $\nu$ is the pulsar's spin frequency and $\dot{\nu}$ is its spin-down rate. To account for changes in the observational backend systems, we separately modeled the white noise parameters EFAC and EQUAD for data collected before and after 2010. The final residuals, dominated by Gaussian random noise, demonstrate that the noise model was effectively characterized.

%This study is limited by the lack of multi-frequency observations, which prevented us from disentangling dispersion measure variations from intrinsic red noise contributions. Future research involving multi-band observations and additional timing data could further improve our understanding of timing noise and its underlying physical processes.  
This analysis of PSR J1741$-$3016 not only demonstrates the capability of \textsc{PINT} to handle time noise in young pulsars but also highlights its potential for study timing noise in large pulsar samples. This will enable a more comprehensive understanding of timing irregularities and their physical origins. Multi-frequency observations and longer time spans will be essential for disentangling contributions from dispersion measure variations and intrinsic red noise. 
%As pulsar timing datasets continue to grow, \textsc{PINT}'s scalability and precision make it a powerful tool for unlocking new insights into pulsar dynamics and timing noise characterization on a population level.  

The computational efficiency metrics in Table \ref{tab:5} demonstrate PINT's practical advantages for long-term timing studies. The 60\% reduction in runtime compared to TEMPONEST, combined with significantly lower CPU utilization (4\% vs. 34\%), makes PINT particularly suitable for analyzing large pulsar populations or performing iterative model refinement. These benchmarks were conducted on enterprise-grade hardware (see Table \ref{tab:5}) 
 to ensure method reproducibility. Notably, the performance advantage persists even when utilizing high-performance computing resources, as PINT's algorithm design minimizes both computational complexity and memory footprint.

\vspace{6pt}

\authorcontributions{onceptualization, J.B. Wang; methodology, J.B. Wang and Y.R. Wen; software, Y.R. Wen; validation, Y.R. Wen; formal analysis, J.B. Wang and W.M. Yan; investigation, J.B. Wang; resources, J.P. Yuan and N. Wang; data curation, J.P. Yuan and Y.R. Wen; writing---original draft preparation, Y.R. Wen and J.B. Wang; writing---review and editing, Y.R. Wen, J.B. Wang, L. Zou and Y. Xia; visualization, Y.R. Wen; supervision, J.B. Wang and W.M. Yan; project administration, J.B. Wang; funding acquisition, J.B. Wang and W.M. Yan. All authors have read and agreed to the published version of the manuscript.}

\funding {This research was funded by the following grants:  Major Science and Technology Program of Xinjiang Uygur Autonomous Region, grant number 2022A03013-4 (to J.B.W.);  Zhejiang Provincial Natural Science Foundation, grant number LY23A030001 (to J.B.W.);  Natural Science Foundation of Xinjiang Uygur Autonomous Region, grant number 2022D01D85 (to J.B.W.);  National Natural Science Foundation of China, grant numbers 12041304 (to J.B.W.), 12273100, 12041303 (to W.M.Y.), and 2020SKA0120200 (to W.M.Y.);  West Light Foundation of Chinese Academy of Sciences, grant number WLFC 2021-XBQNXZ-027 (to W.M.Y.);  National Key R\&D Program of China, grant number 2022YFC2205201 (to W.M.Y.);  Tianshan Talents Program, grant number 2023TSYCTD0013 (to J.P.Y.).  The APC was funded by the National Natural Science Foundation of China (No. 12041304).} 

\dataavailability{The data used in this study were obtained using the Nanshan Radio Telescope of the Xinjiang Astronomical Observatory, Chinese Academy of Sciences. Data can be made available upon reasonable request to the corresponding author.}

\acknowledgments{We acknowledge computational support from the High-Performance Computing Cluster of Xinjiang Astronomical Observatories. Critical open-source tools enabled this work: the PINT team’s frequentist framework; TEMPONEST and ENTERPRISE for Bayesian comparisons; and PSRCHIVE for data standardization. We thank the Nanshan 26 m telescope technical team for sustained observations and the pulsar community for insights shared through IPTA forums.} 

\conflictsofinterest{The authors declare no conflicts of interest. The funders had no role in the design of the study; in the collection, analyses, or interpretation of data; in the writing of the manuscript; or in the decision to publish the results.}

\begin{adjustwidth}{-\extralength}{0cm}
\reftitle{References}

\PublishersNote{}
\end{adjustwidth}

%} % If the paper is ``preprints'', please uncomment this parenthesis.

\begin{thebibliography}{999}

\bibitem[{Manchester}(2017)]{2017JApA...38...42M}
{Manchester}, R.N.
\newblock {Millisecond Pulsars, their Evolution and Applications}.
\newblock {\em J. Astrophys. Astron.} {\bf 2017}, {\em 38},~42.
\newblock {\url{https://doi.org/10.1007/s12036-017-9469-2}}.

\bibitem[{Cromartie} et~al.(2020){Cromartie}, {Fonseca}, {Ransom}, {Demorest},
  {Arzoumanian}, {Blumer}, {Brook}, {DeCesar}, {Dolch}, {Ellis}, {Ferdman},
  {Ferrara}, {Garver-Daniels}, {Gentile}, {Jones}, {Lam}, {Lorimer}, {Lynch},
  {McLaughlin}, {Ng}, {Nice}, {Pennucci}, {Spiewak}, {Stairs}, {Stovall},
  {Swiggum}, and {Zhu}]{2020NatAs...4...72C}
{Cromartie}, H.T.; {Fonseca}, E.; {Ransom}, S.M.; {Demorest}, P.B.;
  {Arzoumanian}, Z.; {Blumer}, H.; {Brook}, P.R.; {DeCesar}, M.E.; {Dolch}, T.;
  {Ellis}, J.A.;  et~al.
\newblock {Relativistic Shapiro delay measurements of an extremely massive
  millisecond pulsar}.
\newblock {\em Nat. Astron.} {\bf 2020}, {\em 4},~72--76.
\newblock {\url{https://doi.org/10.1038/s41550-019-0880-2}}.

\bibitem[{Wolszczan} and {Frail}(1992)]{1992Natur.355..145W}
{Wolszczan}, A.; {Frail}, D.A.
\newblock {A planetary system around the millisecond pulsar PSR1257 + 12}.
\newblock {\em Nature} {\bf 1992}, {\em 355},~145--147.
\newblock {\url{https://doi.org/10.1038/355145a0}}.

\bibitem[{Kramer} et~al.(2021){Kramer}, {Stairs}, {Manchester}, {Wex},
  {Deller}, {Coles}, {Ali}, {Burgay}, {Camilo}, {Cognard}, {Damour},
  {Desvignes}, {Ferdman}, {Freire}, {Grondin}, {Guillemot}, {Hobbs}, {Janssen},
  {Karuppusamy}, {Lorimer}, {Lyne}, {McKee}, {McLaughlin}, {M{\"u}nch},
  {Perera}, {Pol}, {Possenti}, {Sarkissian}, {Stappers}, and
  {Theureau}]{2021PhRvX..11d1050K}
{Kramer}, M.; {Stairs}, I.H.; {Manchester}, R.N.; {Wex}, N.; {Deller}, A.T.;
  {Coles}, W.A.; {Ali}, M.; {Burgay}, M.; {Camilo}, F.; {Cognard}, I.;  et~al.
\newblock {Strong-Field Gravity Tests with the Double Pulsar}.
\newblock {\em Phys. Rev. X} {\bf 2021}, {\em 11},~041050.
\newblock {\url{https://doi.org/10.1103/PhysRevX.11.041050}}.

\bibitem[{Donner} et~al.(2020){Donner}, {Verbiest}, {Tiburzi}, {Os{\l}owski},
  {K{\"u}nsem{\"o}ller}, {Bak Nielsen}, {Grie{\ss}meier}, {Serylak}, {Kramer},
  {Anderson}, {Wucknitz}, {Keane}, {Kondratiev}, {Sobey}, {McKee}, {Bilous},
  {Breton}, {Br{\"u}ggen}, {Ciardi}, {Hoeft}, {van Leeuwen}, and
  {Vocks}]{2020A&A...644A.153D}
{Donner}, J.Y.; {Verbiest}, J.P.W.; {Tiburzi}, C.; {Os{\l}owski}, S.;
  {K{\"u}nsem{\"o}ller}, J.; {Bak Nielsen}, A.S.; {Grie{\ss}meier}, J.M.;
  {Serylak}, M.; {Kramer}, M.; {Anderson}, J.M.;  et~al.
\newblock {Dispersion measure variability for 36 millisecond pulsars at 150 MHz
  with LOFAR}.
\newblock {\em Astron. Astrophys.} {\bf 2020}, {\em 644},~A153.
\newblock {\url{https://doi.org/10.1051/0004-6361/202039517}}.

\bibitem[{Tiburzi} et~al.(2021){Tiburzi}, {Shaifullah}, {Bassa}, {Zucca},
  {Verbiest}, {Porayko}, {van der Wateren}, {Fallows}, {Main}, {Janssen},
  {Anderson}, {Bak Nielsen}, {Donner}, {Keane}, {K{\"u}nsem{\"o}ller},
  {Os{\l}owski}, {Grie{\ss}meier}, {Serylak}, {Br{\"u}ggen}, {Ciardi},
  {Dettmar}, {Hoeft}, {Kramer}, {Mann}, and {Vocks}]{2021A&A...647A..84T}
{Tiburzi}, C.; {Shaifullah}, G.M.; {Bassa}, C.G.; {Zucca}, P.; {Verbiest},
  J.P.W.; {Porayko}, N.K.; {van der Wateren}, E.; {Fallows}, R.A.; {Main},
  R.A.; {Janssen}, G.H.;  et~al.
\newblock {The impact of solar wind variability on pulsar timing}.
\newblock {\em Astron. Astrophys.} {\bf 2021}, {\em 647},~A84.
\newblock {\url{https://doi.org/10.1051/0004-6361/202039846}}.

\bibitem[{Hobbs} et~al.(2020){Hobbs}, {Guo}, {Caballero}, {Coles}, {Lee},
  {Manchester}, {Reardon}, {Matsakis}, {Tong}, {Arzoumanian}, {Bailes},
  {Bassa}, {Bhat}, {Brazier}, {Burke-Spolaor}, {Champion}, {Chatterjee},
  {Cognard}, {Dai}, {Desvignes}, {Dolch}, {Ferdman}, {Graikou}, {Guillemot},
  {Janssen}, {Keith}, {Kerr}, {Kramer}, {Lam}, {Liu}, {Lyne}, {Lazio}, {Lynch},
  {McKee}, {McLaughlin}, {Mingarelli}, {Nice}, {Os{\l}owski}, {Pennucci},
  {Perera}, {Perrodin}, {Possenti}, {Russell}, {Sanidas}, {Sesana},
  {Shaifullah}, {Shannon}, {Simon}, {Spiewak}, {Stairs}, {Stappers}, {Swiggum},
  {Taylor}, {Theureau}, {Toomey}, {van Haasteren}, {Wang}, {Wang}, and
  {Zhu}]{2020MNRAS.491.5951H}
{Hobbs}, G.; {Guo}, L.; {Caballero}, R.N.; {Coles}, W.; {Lee}, K.J.;
  {Manchester}, R.N.; {Reardon}, D.J.; {Matsakis}, D.; {Tong}, M.L.;
  {Arzoumanian}, Z.;  et~al.
\newblock {A pulsar-based time-scale from the International Pulsar Timing
  Array}.
\newblock {\em Mon. Not. R. Astron. Soc.} {\bf 2020}, {\em 491},~5951--5965.
\newblock {\url{https://doi.org/10.1093/mnras/stz3071}}.

\bibitem[{Caballero} et~al.(2018){Caballero}, {Guo}, {Lee}, {Lazarus},
  {Champion}, {Desvignes}, {Kramer}, {Plant}, {Arzoumanian}, {Bailes}, {Bassa},
  {Bhat}, {Brazier}, {Burgay}, {Burke-Spolaor}, {Chamberlin}, {Chatterjee},
  {Cognard}, {Cordes}, {Dai}, {Demorest}, {Dolch}, {Ferdman}, {Fonseca},
  {Gair}, {Garver-Daniels}, {Gentile}, {Gonzalez}, {Graikou}, {Guillemot},
  {Hobbs}, {Janssen}, {Karuppusamy}, {Keith}, {Kerr}, {Lam}, {Lasky}, {Lazio},
  {Levin}, {Liu}, {Lommen}, {Lorimer}, {Lynch}, {Madison}, {Manchester},
  {McKee}, {McLaughlin}, {McWilliams}, {Mingarelli}, {Nice}, {Os{\l}owski},
  {Palliyaguru}, {Pennucci}, {Perera}, {Perrodin}, {Possenti}, {Ransom},
  {Reardon}, {Sanidas}, {Sesana}, {Shaifullah}, {Shannon}, {Siemens}, {Simon},
  {Spiewak}, {Stairs}, {Stappers}, {Stinebring}, {Stovall}, {Swiggum},
  {Taylor}, {Theureau}, {Tiburzi}, {Toomey}, {van Haasteren}, {van Straten},
  {Verbiest}, {Wang}, {Zhu}, and {Zhu}]{2018MNRAS.481.5501C}
{Caballero}, R.N.; {Guo}, Y.J.; {Lee}, K.J.; {Lazarus}, P.; {Champion}, D.J.;
  {Desvignes}, G.; {Kramer}, M.; {Plant}, K.; {Arzoumanian}, Z.; {Bailes}, M.;
  et~al.
\newblock {Studying the Solar system with the International Pulsar Timing
  Array}.
\newblock {\em Mon. Not. R. Astron. Soc.} {\bf 2018}, {\em 481},~5501--5516.
\newblock {\url{https://doi.org/10.1093/mnras/sty2632}}.

\bibitem[{Xu} et~al.(2023){Xu}, {Chen}, {Guo}, {Jiang}, {Wang}, {Xu}, {Xue},
  {Nicolas Caballero}, {Yuan}, {Xu}, {Wang}, {Hao}, {Luo}, {Lee}, {Han},
  {Jiang}, {Shen}, {Wang}, {Wang}, {Xu}, {Wu}, {Manchester}, {Qian}, {Guan},
  {Huang}, {Sun}, and {Zhu}]{2023RAA....23g5024X}
{Xu}, H.; {Chen}, S.; {Guo}, Y.; {Jiang}, J.; {Wang}, B.; {Xu}, J.; {Xue}, Z.;
  {Nicolas Caballero}, R.; {Yuan}, J.; {Xu}, Y.;  et~al.
\newblock {Searching for the Nano-Hertz Stochastic Gravitational Wave
  Background with the Chinese Pulsar Timing Array Data Release I}.
\newblock {\em Res. Astron. Astrophys.} {\bf 2023}, {\em
  23},~075024.
\newblock {\url{https://doi.org/10.1088/1674-4527/acdfa5}}.

\bibitem[Foster and Backer(1990)]{Foster1990}
Foster, R.S.; Backer, D.C.
\newblock Constructing a Pulsar Timing Array.
\newblock {\em  Astrophys. J.} {\bf 1990}, {\em 361},~300.
\newblock {\url{https://doi.org/10.1086/169195}}.

\bibitem[Staelin and Reifenstein(1968)]{staelin1968}
Staelin, D.H.; Reifenstein, E.C.
\newblock Pulsating {{Radio Sources}} near the {{Crab Nebula}}.
\newblock {\em Science} {\bf 1968}, {\em 162},~1481--1483.

\bibitem[{Keith} et~al.(2013){Keith}, {Coles}, {Shannon}, {Hobbs},
  {Manchester}, {Bailes}, {Bhat}, {Burke-Spolaor}, {Champion}, {Chaudhary},
  {Hotan}, {Khoo}, {Kocz}, {Os{\l}owski}, {Ravi}, {Reynolds}, {Sarkissian},
  {van Straten}, and {Yardley}]{2013MNRAS.429.2161K}
{Keith}, M.J.; {Coles}, W.; {Shannon}, R.M.; {Hobbs}, G.B.; {Manchester}, R.N.;
  {Bailes}, M.; {Bhat}, N.D.R.; {Burke-Spolaor}, S.; {Champion}, D.J.;
  {Chaudhary}, A.;  et~al.
\newblock {Measurement and correction of variations in interstellar dispersion
  in high-precision pulsar timing}.
\newblock {\em Mon. Not. R. Astron. Soc.} {\bf 2013}, {\em 429},~2161--2174.
\newblock {\url{https://doi.org/10.1093/mnras/sts486}}.

\bibitem[{Hellings} and {Downs}(1983)]{1983ApJ...265L..39H}
{Hellings}, R.W.; {Downs}, G.S.
\newblock {Upper limits on the isotropic gravitational radiation background
  from pulsar timing analysis.}
\newblock {\em  Astrophys. J. Lett.} {\bf 1983}, {\em 265},~L39--L42.
\newblock {\url{https://doi.org/10.1086/183954}}.

\bibitem[{Baym} et~al.(1969){Baym}, {Pethick}, {Pines}, and
  {Ruderman}]{1969Natur.224..872B}
{Baym}, G.; {Pethick}, C.; {Pines}, D.; {Ruderman}, M.
\newblock {Spin Up in Neutron Stars : The Future of the Vela Pulsar}.
\newblock {\em Nature} {\bf 1969}, {\em 224},~872--874.
\newblock {\url{https://doi.org/10.1038/224872a0}}.

\bibitem[{Anderson} and {Itoh}(1975)]{1975Natur.256...25A}
{Anderson}, P.W.; {Itoh}, N.
\newblock {Pulsar glitches and restlessness as a hard superfluidity
  phenomenon}.
\newblock {\em Nature} {\bf 1975}, {\em 256},~25--27.
\newblock {\url{https://doi.org/10.1038/256025a0}}.

\bibitem[{Yuan} et~al.(2010){Yuan}, {Wang}, {Manchester}, and {Liu}]{yuan2010}
{Yuan}, J.P.; {Wang}, N.; {Manchester}, R.N.; {Liu}, Z.Y.
\newblock {29 glitches detected at Urumqi Observatory}.
\newblock {\em Mon. Not. R. Astron. Soc.} {\bf 2010}, {\em 404},~289--304.
\newblock {\url{https://doi.org/10.1111/j.1365-2966.2010.16272.x}}.

\bibitem[Hobbs et~al.(2010)Hobbs, Lyne, and Kramer]{hobbs2010}
Hobbs, G.; Lyne, A.G.; Kramer, M.
\newblock An Analysis of the Timing Irregularities for 366 Pulsars.
\newblock {\em Mon. Not. R. Astron. Soc.} {\bf 2010},
  {\em 402},~1027--1048.
\newblock {\url{https://doi.org/10.1111/j.1365-2966.2009.15938.x}}.

\bibitem[{Parthasarathy} et~al.(2019){Parthasarathy}, {Shannon}, {Johnston},
  {Lentati}, {Bailes}, {Dai}, {Kerr}, {Manchester}, {Os{\l}owski}, {Sobey},
  {van Straten}, and {Weltevrede}]{2019MNRAS.489.3810P}
{Parthasarathy}, A.; {Shannon}, R.M.; {Johnston}, S.; {Lentati}, L.; {Bailes},
  M.; {Dai}, S.; {Kerr}, M.; {Manchester}, R.N.; {Os{\l}owski}, S.; {Sobey},
  C.;  et~al.
\newblock {Timing of young radio pulsars - I. Timing noise, periodic
  modulation, and proper motion}.
\newblock {\em Mon. Not. R. Astron. Soc.} {\bf 2019}, {\em 489},~3810--3826.
\newblock {\url{https://doi.org/10.1093/mnras/stz2383}}.

\bibitem[{Blandford} et~al.(1984){Blandford}, {Narayan}, and
  {Romani}]{1984JApA....5..369B}
{Blandford}, R.; {Narayan}, R.; {Romani}, R.W.
\newblock {Arrival Time Analysis for a Millisecond Pulsar}.
\newblock {\em J. Astrophys. Astron.} {\bf 1984}, {\em
  5},~369--388.
\newblock {\url{https://doi.org/10.1007/BF02714466}}.

\bibitem[{Rawley} et~al.(1988){Rawley}, {Taylor}, and
  {Davis}]{1988ApJ...326..947R}
{Rawley}, L.A.; {Taylor}, J.H.; {Davis}, M.M.
\newblock {Fundamental Astrometry and Millisecond Pulsars}.
\newblock {\em  Astrophys. J.} {\bf 1988}, {\em 326},~947.
\newblock {\url{https://doi.org/10.1086/166153}}.

\bibitem[{Cordes} and {Greenstein}(1981)]{1981ApJ...245.1060C}
{Cordes}, J.M.; {Greenstein}, G.
\newblock {Pulsar timing .IV. Physical models for timing noise processes.}
\newblock {\em  Astrophys. J.} {\bf 1981}, {\em 245},~1060--1079.
\newblock {\url{https://doi.org/10.1086/158883}}.

\bibitem[Nice et~al.(2015)Nice, Demorest, Stairs, Manchester, Taylor, Peters,
  Weisberg, Irwin, Wex, and Huang]{2015ascl.soft09002N}
Nice, D.; Demorest, P.; Stairs, I.; Manchester, R.; Taylor, J.; Peters, W.;
  Weisberg, J.; Irwin, A.; Wex, N.; Huang, Y.
\newblock Tempo: Pulsar timing data analysis.
\newblock Astrophysics Source Code Library, record ascl:1509.002,  2015.
\newblock Online resource: \url{https://tempo.sourceforge.net/}.
\newblock Software version: 2015 release. 


\bibitem[Hobbs et~al.(2006)Hobbs, Edwards, and Manchester]{hobbs2006}
Hobbs, G.B.; Edwards, R.T.; Manchester, R.N.
\newblock Tempo2, a New Pulsar-Timing Package - {{I}}. {{An}} Overview: Tempo2,
  a New Pulsar-Timing Package - {{I}}. {{Overview}}.
\newblock {\em Mon. Not. R. Astron. Soc.} {\bf 2006},
  {\em 369},~655--672.
\newblock {\url{https://doi.org/10.1111/j.1365-2966.2006.10302.x}}.

\bibitem[Luo et~al.(2021)Luo, Ransom, Demorest, Ray, Archibald, Kerr, Jennings,
  Bachetti, {van Haasteren}, Champagne, Colen, Phillips, Zimmerman, Stovall,
  Lam, and Jenet]{luo2021}
Luo, J.; Ransom, S.; Demorest, P.; Ray, P.S.; Archibald, A.; Kerr, M.;
  Jennings, R.J.; Bachetti, M.; {van Haasteren}, R.; Champagne, C.A.;  et~al.
\newblock {{PINT}}: {{A Modern Software Package}} for {{Pulsar Timing}}.
\newblock {\em  Astrophys. J.} {\bf 2021}, {\em 911},~45.
\newblock {\url{https://doi.org/10.3847/1538-4357/abe62f}}.

\bibitem[{Johnson} et~al.(2024){Johnson}, {Meyers}, {Baker}, {Cornish},
  {Hazboun}, {Littenberg}, {Romano}, {Taylor}, {Vallisneri}, {Vigeland},
  {Olum}, {Siemens}, {Ellis}, {van Haasteren}, {Hourihane}, {Agazie},
  {Anumarlapudi}, {Archibald}, {Arzoumanian}, {Blecha}, {Brazier}, {Brook},
  {Burke-Spolaor}, {B{\'e}csy}, {Casey-Clyde}, {Charisi}, {Chatterjee},
  {Chatziioannou}, {Cohen}, {Cordes}, {Crawford}, {Cromartie}, {Crowter},
  {Decesar}, {Demorest}, {Dolch}, {Drachler}, {Ferrara}, {Fiore}, {Fonseca},
  {Freedman}, {Garver-Daniels}, {Gentile}, {Glaser}, {Good}, {G{\"u}ltekin},
  {Jennings}, {Jones}, {Kaiser}, {Kaplan}, {Kelley}, {Kerr}, {Key}, {Laal},
  {Lam}, {Lamb}, {Lazio}, {Lewandowska}, {Liu}, {Lorimer}, {Luo}, {Lynch},
  {Ma}, {Madison}, {McEwen}, {McKee}, {McLaughlin}, {McMann}, {Meyers},
  {Mingarelli}, {Mitridate}, {Ng}, {Nice}, {Ocker}, {Pennucci}, {Perera},
  {Pol}, {Radovan}, {Ransom}, {Ray}, {Sardesai}, {Schmiedekamp},
  {Schmiedekamp}, {Schmitz}, {Shapiro-Albert}, {Simon}, {Siwek}, {Stairs},
  {Stinebring}, {Stovall}, {Susobhanan}, {Swiggum}, {Turner}, {Unal}, {Wahl},
  {Witt}, {Young}, and {Nanograv Collaboration}]{2024PhRvD.109j3012J}
{Johnson}, A.D.; {Meyers}, P.M.; {Baker}, P.T.; {Cornish}, N.J.; {Hazboun},
  J.S.; {Littenberg}, T.B.; {Romano}, J.D.; {Taylor}, S.R.; {Vallisneri}, M.;
  {Vigeland}, S.J.;  et~al.
\newblock {NANOGrav 15-year gravitational-wave background methods}.
\newblock {\em \prd} {\bf 2024}, {\em 109},~103012.
\newblock {\url{https://doi.org/10.1103/PhysRevD.109.103012}}.

\bibitem[Lentati et~al.(2014)Lentati, Alexander, Hobson, Feroz, {van
  Haasteren}, Lee, and Shannon]{lentati2014}
Lentati, L.; Alexander, P.; Hobson, M.P.; Feroz, F.; {van Haasteren}, R.; Lee,
  K.; Shannon, R.M.
\newblock {{TempoNest}}: {{A Bayesian}} Approach to Pulsar Timing Analysis.
\newblock {\em Mon. Not. R. Astron. Soc.} {\bf 2014},
  {\em 437},~3004--3023.
\newblock {\url{https://doi.org/10.1093/mnras/stt2122}}.

\bibitem[{Agazie} et~al.(2023){Agazie}, {Anumarlapudi}, {Archibald}, {Baker},
  {B{\'e}csy}, {Blecha}, {Bonilla}, {Brazier}, {Brook}, {Burke-Spolaor},
  {Burnette}, {Case}, {Casey-Clyde}, {Charisi}, {Chatterjee}, {Chatziioannou},
  {Cheeseboro}, {Chen}, {Cohen}, {Cordes}, {Cornish}, {Crawford}, {Cromartie},
  {Crowter}, {Cutler}, {D'Orazio}, {Decesar}, {Degan}, {Demorest}, {Deng},
  {Dolch}, {Drachler}, {Ferrara}, {Fiore}, {Fonseca}, {Freedman}, {Gardiner},
  {Garver-Daniels}, {Gentile}, {Gersbach}, {Glaser}, {Good}, {G{\"u}ltekin},
  {Hazboun}, {Hourihane}, {Islo}, {Jennings}, {Johnson}, {Jones}, {Kaiser},
  {Kaplan}, {Kelley}, {Kerr}, {Key}, {Laal}, {Lam}, {Lamb}, {Lazio},
  {Lewandowska}, {Littenberg}, {Liu}, {Luo}, {Lynch}, {Ma}, {Madison},
  {McEwen}, {McKee}, {McLaughlin}, {McMann}, {Meyers}, {Meyers}, {Mingarelli},
  {Mitridate}, {Natarajan}, {Ng}, {Nice}, {Ocker}, {Olum}, {Pennucci},
  {Perera}, {Petrov}, {Pol}, {Radovan}, {Ransom}, {Ray}, {Romano}, {Runnoe},
  {Sardesai}, {Schmiedekamp}, {Schmiedekamp}, {Schmitz}, {Schult},
  {Shapiro-Albert}, {Siemens}, {Simon}, {Siwek}, {Stairs}, {Stinebring},
  {Stovall}, {Sun}, {Susobhanan}, {Swiggum}, {Taylor}, {Taylor}, {Turner},
  {Unal}, {Vallisneri}, {Vigeland}, {Wachter}, {Wahl}, {Wang}, {Witt},
  {Wright}, {Young}, and {Nanograv Collaboration}]{2023ApJ...952L..37A}
{Agazie}, G.; {Anumarlapudi}, A.; {Archibald}, A.M.; {Baker}, P.T.;
  {B{\'e}csy}, B.; {Blecha}, L.; {Bonilla}, A.; {Brazier}, A.; {Brook}, P.R.;
  {Burke-Spolaor}, S.;  et~al.
\newblock {The NANOGrav 15 yr Data Set: Constraints on Supermassive Black Hole
  Binaries from the Gravitational-wave Background}.
\newblock {\em  Astrophys. J. Lett.} {\bf 2023}, {\em 952},~L37.
\newblock {\url{https://doi.org/10.3847/2041-8213/ace18b}}.

\bibitem[{Reardon} et~al.(2023){Reardon}, {Zic}, {Shannon}, {Hobbs}, {Bailes},
  {Di Marco}, {Kapur}, {Rogers}, {Thrane}, {Askew}, {Bhat}, {Cameron},
  {Cury{\l}o}, {Coles}, {Dai}, {Goncharov}, {Kerr}, {Kulkarni}, {Levin},
  {Lower}, {Manchester}, {Mandow}, {Miles}, {Nathan}, {Os{\l}owski}, {Russell},
  {Spiewak}, {Zhang}, and {Zhu}]{2023ApJ...951L...6R}
{Reardon}, D.J.; {Zic}, A.; {Shannon}, R.M.; {Hobbs}, G.B.; {Bailes}, M.; {Di
  Marco}, V.; {Kapur}, A.; {Rogers}, A.F.; {Thrane}, E.; {Askew}, J.;  et~al.
\newblock {Search for an Isotropic Gravitational-wave Background with the
  Parkes Pulsar Timing Array}.
\newblock {\em  Astrophys. J. Lett.} {\bf 2023}, {\em 951},~L6.
\newblock {\url{https://doi.org/10.3847/2041-8213/acdd02}}.

\bibitem[{Demorest} et~al.(2013){Demorest}, {Ferdman}, {Gonzalez}, {Nice},
  {Ransom}, {Stairs}, {Arzoumanian}, {Brazier}, {Burke-Spolaor}, {Chamberlin},
  {Cordes}, {Ellis}, {Finn}, {Freire}, {Giampanis}, {Jenet}, {Kaspi}, {Lazio},
  {Lommen}, {McLaughlin}, {Palliyaguru}, {Perrodin}, {Shannon}, {Siemens},
  {Stinebring}, {Swiggum}, and {Zhu}]{2013ApJ...762...94D}
{Demorest}, P.B.; {Ferdman}, R.D.; {Gonzalez}, M.E.; {Nice}, D.; {Ransom}, S.;
  {Stairs}, I.H.; {Arzoumanian}, Z.; {Brazier}, A.; {Burke-Spolaor}, S.;
  {Chamberlin}, S.J.;  et~al.
\newblock {Limits on the Stochastic Gravitational Wave Background from the
  North American Nanohertz Observatory for Gravitational Waves}.
\newblock {\em  Astrophys. J.} {\bf 2013}, {\em 762},~94.
\newblock {\url{https://doi.org/10.1088/0004-637X/762/2/94}}.

\bibitem[Burnham and Anderson(2004)]{burnham2004}
Burnham, K.P.; Anderson, D.R.
\newblock Multimodel {{Inference}}: {{Understanding AIC}} and {{BIC}} in
  {{Model Selection}}.
\newblock {\em Sociol. Methods Res.} {\bf 2004}, {\em
  33},~261--304.
\newblock {\url{https://doi.org/10.1177/0049124104268644}}.

\bibitem[{Jones} and {Qin}(2022)]{2022AnRSA...9..557J}
{Jones}, G.L.; {Qin}, Q.
\newblock {Markov Chain Monte Carlo in Practice}.
\newblock {\em Annu. Rev. Stat. Its Appl.} {\bf 2022},
  {\em 9},~557--578.
\newblock {\url{https://doi.org/10.1146/annurev-statistics-040220-090158}}.

\bibitem[Wang et~al.(2001)Wang, Manchester, Zhang, Wu, Yusup, Lyne, Cheng, and
  Chen]{wang2001}
Wang, N.; Manchester, R.; Zhang, J.; Wu, X.; Yusup, A.; Lyne, A.; Cheng, K.;
  Chen, M.
\newblock Pulsar Timing at {{Urumqi Astronomical Observatory}}: Observing
  System and Results.
\newblock {\em Mon. Not. R. Astron. Soc.} {\bf 2001},
  {\em 328},~855--866.
\newblock {\url{https://doi.org/10.1046/j.1365-8711.2001.04926.x}}.

\bibitem[Manchester et~al.(2013)Manchester, Hobbs, Bailes, Coles, Van~Straten,
  Keith, Shannon, Bhat, Brown, {Burke-Spolaor}, Champion, Chaudhary, Edwards,
  Hampson, Hotan, Jameson, Jenet, Kesteven, Khoo, Kocz, Maciesiak, Oslowski,
  Ravi, Reynolds, Sarkissian, Verbiest, Wen, Wilson, Yardley, Yan, and
  You]{manchester2013}
Manchester, R.N.; Hobbs, G.; Bailes, M.; Coles, W.A.; Van~Straten, W.; Keith,
  M.J.; Shannon, R.M.; Bhat, N.D.R.; Brown, A.; {Burke-Spolaor}, S.G.;  et~al.
\newblock The {{Parkes Pulsar Timing Array Project}}.
\newblock {\em Publ. Astron. Soc. Aust.} {\bf
  2013}, {\em 30},~e017.
\newblock {\url{https://doi.org/10.1017/pasa.2012.017}}.

\bibitem[Yuan et~al.(2017)Yuan, Manchester, Wang, Wang, Zhou, Yan, and
  Liu]{yuan2017}
Yuan, J.P.; Manchester, R.N.; Wang, N.; Wang, J.B.; Zhou, X.; Yan, W.M.; Liu,
  Z.Y.
\newblock Pulse Profiles and Timing of {{PSR J1757}}\$-\$2421.
\newblock {\em Mon. Not. R. Astron. Soc.} {\bf 2017},
  {\em 466},~1234--1241.
\newblock {\url{https://doi.org/10.1093/mnras/stw3203}}.

\bibitem[Morris et~al.(2002)Morris, Hobbs, Lyne, Stairs, Camilo, Manchester,
  Possenti, Bell, Kaspi, D'Amico, McKay, Crawford, and
  Kramer]{morrisParkesMultibeamPulsar2002}
Morris, D.J.; Hobbs, G.; Lyne, A.G.; Stairs, I.H.; Camilo, F.; Manchester,
  R.N.; Possenti, A.; Bell, J.F.; Kaspi, V.M.; D'Amico, N.;  et~al.
\newblock The {{Parkes Multibeam Pulsar Survey}}--{{II}}. {{Discovery}} and
  {{Timing}} of 120 {{Pulsars}}.
\newblock {\em Mon. Not. R. Astron. Soc.} {\bf 2002},
  {\em 335},~275--290.
\newblock {\url{https://doi.org/10.1046/j.1365-8711.2002.05551.x}}.

\bibitem[{Ro{\.z}ko} et~al.(2021){Ro{\.z}ko}, {Basu}, {Kijak}, and
  {Lewandowski}]{2021ApJ...922..125R}
{Ro{\.z}ko}, K.; {Basu}, R.; {Kijak}, J.; {Lewandowski}, W.
\newblock {The uGMRT Observations of Three New Gigahertz-peaked Spectra
  Pulsars}.
\newblock {\em  Astrophys. J.} {\bf 2021}, {\em 922},~125.
\newblock {\url{https://doi.org/10.3847/1538-4357/ac23dc}}.

\bibitem[Hotan et~al.(2004)Hotan, Van~Straten, and Manchester]{hotan2004}
Hotan, A.W.; Van~Straten, W.; Manchester, R.N.
\newblock {\textsc{Psrchive}} and {\textsc{Psrfits}} : {{An Open Approach}} to
  {{Radio Pulsar Data Storage}} and {{Analysis}}.
\newblock {\em Publ. Astron. Soc. Aust.} {\bf
  2004}, {\em 21},~302--309.
\newblock {\url{https://doi.org/10.1071/AS04022}}.

\bibitem[{van Straten} et~al.(2012){van Straten}, {Demorest}, and
  {Oslowski}]{straten2012}
{van Straten}, W.; {Demorest}, P.; {Oslowski}, S.
\newblock {Pulsar Data Analysis with PSRCHIVE}.  	\emph{arXiv} \textbf{2012}.
\newblock {\url{https://doi.org/10.48550/arXiv.1205.6276}}.

\bibitem[{Manchester} et~al.(2005){Manchester}, {Hobbs}, {Teoh}, and
  {Hobbs}]{2005AJ....129.1993M}
{Manchester}, R.N.; {Hobbs}, G.B.; {Teoh}, A.; {Hobbs}, M.
\newblock {The Australia Telescope National Facility Pulsar Catalogue}.
\newblock {\em  Astron. J.} {\bf 2005}, {\em 129},~1993--2006.
\newblock {\url{https://doi.org/10.1086/428488}}.

\bibitem[Susobhanan et~al.(2024)Susobhanan, Kaplan, Archibald, Luo, Ray,
  Pennucci, Ransom, Agazie, Fiore, Larsen, O'Neill, Van~Haasteren,
  Anumarlapudi, Bachetti, Bhakta, Champagne, Cromartie, Demorest, Jennings,
  Kerr, Levina, McEwen, {Shapiro-Albert}, and Swiggum]{susobhanan2024}
Susobhanan, A.; Kaplan, D.L.; Archibald, A.M.; Luo, J.; Ray, P.S.; Pennucci,
  T.T.; Ransom, S.M.; Agazie, G.; Fiore, W.; Larsen, B.;  et~al.
\newblock {{PINT}}: {{Maximum-likelihood Estimation}} of {{Pulsar Timing Noise
  Parameters}}.
\newblock {\em  Astrophys. J.} {\bf 2024}, {\em 971},~150.
\newblock {\url{https://doi.org/10.3847/1538-4357/ad59f7}}.

\bibitem[{The NANOGrav Collaboration} et~al.(2015){The NANOGrav Collaboration},
  Arzoumanian, Brazier, {Burke-Spolaor}, Chamberlin, Chatterjee, Christy,
  Cordes, Cornish, Crowter, Demorest, Dolch, Ellis, Ferdman, Fonseca,
  {Garver-Daniels}, Gonzalez, Jenet, Jones, Jones, Kaspi, Koop, Lam, Lazio,
  Levin, Lommen, Lorimer, Luo, Lynch, Madison, McLaughlin, McWilliams, Nice,
  Palliyaguru, Pennucci, Ransom, Siemens, Stairs, Stinebring, Stovall, Swiggum,
  Vallisneri, Haasteren, Wang, and Zhu]{thenanogravcollaboration2015}
{The NANOGrav Collaboration}; Arzoumanian, Z.; Brazier, A.; {Burke-Spolaor},
  S.; Chamberlin, S.; Chatterjee, S.; Christy, B.; Cordes, J.M.; Cornish, N.;
  Crowter, K.;  et~al.
\newblock {{The NANOGrav 
 nine-year data set}}: {{observations}}, {{arrival time
  measurements}}, {{and analysis of}} 37 {{millisecond pulsars}}.
\newblock {\em  Astrophys. J.} {\bf 2015}, {\em 813},~65.
\newblock {\url{https://doi.org/10.1088/0004-637X/813/1/65}}.

\end{thebibliography}
\end{document}